# 10 to 50 nm Long Quasi Ballistic Carbon Nanotube Devices Obtained Without Complex Lithography

Ali Javey, Pengfei Qi, Qian Wang and Hongjie Dai*

Department of chemistry and Laboratory of Advanced Materials, Stanford University, CA 94305, USA

## Abstract

A simple method combining photolithography and shadow (or angle) evaporation is developed to fabricate single-walled carbon nanotube (SWCNT) devices with tube lengths L~10-50 nm between metal contacts. Large numbers of such short devices are obtained without the need of complex tools such as electron beam lithography. Metallic SWCNTs with lengths ~ 10 nm, near the mean free path (mfp) of $l_{op}$~15 nm for optical phonon scattering, exhibit near-ballistic transport at high biases and can carry unprecedented 100 µA currents per tube. Semiconducting SWCNT field-effect transistors (FETs) with ~ 50 nm channel lengths are routinely produced to achieve quasi-ballistic operations for molecular transistors. The results demonstrate highly length-scaled and high-performance interconnects and transistors realized with SWCNTs.

* Correspondence to hdai@stanford.edu

Ballistic transport refers to the motion of charge carriers driven by electric fields in a conducting or semiconducting material without scattering. It is a highly desirable phenomenon for a wide range of applications needing high currents, high speeds and low power dissipations. Single-walled carbon nanotubes have been suggested as candidate materials for future electronics including electrical interconnects and field-effect transistors (1-4), for which high current operations are important. It has been shown that among various scattering mechanisms in SWCNTs, inelastic optical phonon (OP) scattering/emission has the shortest mean free path of $l_{op}$ ~ 15 nm, followed by elastic acoustic phonon scattering mfp of $l_{ap}$ ~300-700 nm at room temperature and defect scattering mfp of $l_e$ ~ 1-3 µm (5-8). Transport can be ballistic in relatively long SWCNTs at low bias voltages and electric fields. In the high bias and current regime, however, back scattering of energetic electrons by OP emission causes large channel resistance and limits the current flow, unless the length of the SWCNTs can be reduced below $l_{op}$ ~ 10-15 nm (5,6,8). There have been several attempts in length scaling of SWCNT devices and the shortest devices thus far are ~ 50 nm long for both metallic (6,8) and semiconducting SWCNTs (9,10) fabricated by electron beam lithography. Defining device lengths below 50 nm is difficult by lithographic techniques, while it remains interesting and important to do so in order to investigate the ultimate current carrying capability of SWCNTs and push the performance limit of molecular transistors.

Here, we show that by using a simple photolithography and shadow evaporation technique, one can readily obtain large numbers of devices comprised of ultra-short SWCNTs down to ~10 nm between two metal contacts. We find that individual ~ 10 nm long metallic SWCNTs can carry ~100 µA per tube and are essentially macromolecules with highly ballistic transport properties. By connecting less than 10 such ultra-short SWCNTs in parallel, one can reach macroscopic current flows on the order of 1 mA. The same fabrication method also affords SWCNT field-effect transistors with ~ 50 nm channel lengths without relying on electron beam lithography. These highly length-scaled FETs can deliver near ballistic currents for transistor operations in the on-state.

We first synthesized SWCNTs by chemical vapor deposition (CVD) of methane (11) from an array of catalytically patterned sites on Si/SiO$_2$ substrates (oxide thickness ~ 10 nm in regions that SWCNTs were grown and ~100 nm in other regions of the substrates). We then formed an array of SWCNT devices, each comprised of two Pd metal contacts (9) spaced at ~3 μm, by photolithography patterning of resist, metal deposition and liftoff. The thickness of this first Pd metal deposition was varied from $t_1$=30 to 50 nm and sometimes a nominally 0.5 nm thick Ti was used as an adhesion layer for Pd. The boron doped Si substrate was used as the gate electrode for the SWCNT devices. Characterization by electrical transport, i.e., conductance vs. gate-voltage (G vs. $V_{GS}$) and atomic force microscopy (AFM) was used to identify devices with individual metallic or semiconducting SWCNTs bridging the electrode pairs in the array. On a typical 4 mm by 4 mm chip, tens of individual SWCNT devices were obtained out of an array of ~100.

With these 'long' (~3 μm) SWCNT devices, we carried out a second photolithography step to open windows (~10 by 10 μm in size) in a photo-resist layer over the SWCNTs and the electrodes, and performed angle electron-beam evaporation of ~ 8 to 20 nm thick Pd followed by lift-off. Due to directional metal deposition of electron-beam evaporation, placing the substrate normal at an angle (θ) to the deposition direction afforded shadow formation next to the pre-formed Pd electrodes. That is, the existing metal electrodes were used as shadow masks for the second metal deposition step to produce small gaps L ~ $t_1 \times tan(\theta)$ ~ 10-50 nm (for $t_1$=30-50 nm and $\theta$~20-45º) between source-drain (S-D) electrodes (Fig.2a). SWCNTs bridging these ultra-small S-D electrodes thus afforded ultra-short tube devices. Fig. 2b shows a scanning electron microscopy (SEM) image of a S/D electrode pair bridged by an L~ 15±5 nm long SWCNT. AFM was also used to characterize these short SWCNT devices. However, for gaps < ~ 30 nm, we found that it is difficult for the AFM tips to reach into the gaps and produce high quality images.

Current saturations are known to occur in long metallic SWCNTs under high bias voltages at the ~ 20-25 µA level (Fig.1, tube length L~ 1 µm) due to optical phonon scattering with a short mfp of $l_{op}$ ~ 10-15 nm (5,6,8). We observe drastically different transport properties for L ~ 15±5 nm metallic SWCNTs (diameter $d$ ~ 2 nm), as can be gleaned from the current ($I_{DS}$) vs. bias voltage ($V_{DS}$) curve in Fig.2c. In strong contrast to micron-long tubes, up to 110 µA of current can be delivered through the ultra-short SWCNT, corresponding to ~ $4 \times 10^9$ A/cm$^2$ current density (or 55,000 A/m, normalized by $d$). This is among the highest current density tolerable by any conductor at room temperature. The ~ 100 µA is the highest current transported through a SWCNT, made possible here simply by forming the shortest and thus most ballistic nantoube channels. In the low bias regime, the $I_{DS}$-$V_{DS}$ curve is linear with a slope of G=$G_0 T_D T_S$, where $G_0 = 4e^2/h$ and $T_S$~$T_D$~0.85 are the transmission probabilities at the S and D Pd contacts respectively. At high biases, optical phonon scattering is the dominant scattering mechanism inside the short nanotube. The transmission probability $T$ due to OP scattering is related to the nanotube length L and $l_{op}$ by $T = l_{op}/(l_{op} + L)$. The conductance of the nanotube device is $G = \Delta I_{DS} / \Delta V_{DS} = G_0 T_S T_D T$. From the slope of the $I_{DS}$-$V_{DS}$ curve in the high bias regime and $T_S \sim T_D \sim 0.85$, we obtain $T \sim 0.4$ due to OP scattering. Since L~15 nm as measured by SEM, a transmission probability of T~0.4 suggests a mfp of $l_{op}$ ~10 nm, which is similar to $l_{op}$ ~10-15 nm measured previously by independent experiments and groups (5,6,8).

Our results show that short SWCNTs are quasi-ballistic macromolecules that can survive high bias voltages and currents. The $I_{DS}$-$V_{DS}$ curve exhibits an upturn in the slope for biases beyond ~1.3 V (Fig.2), attributed to the onset of additional transport through the first non-crossoing sub-band with an energy gap of $\sim 2.6(eV)/d(in\,nm) \approx 1.3 eV$ (6). Note that the $\sim 4 \times 10^9$ A/cm$^2$ current density is three orders of magnitude higher than that tolerable by a typical metal before breakdown via electro-migration. Carbon nanotubes can sustain such high currents due to the strong chemical bonding in the covalent sp$^2$ carbon network. Using the angle evaporation process described above, we have formed eight L ~ 15 nm short devices on an individual SWCNT and then connected them in parallel with comb-like inter-digitized electrodes (Fig. 3). Up to 1 mA current can be flowed through such a device (Fig. 3b). This suggests that if SWCNTs can be close-packed with every tube ohmically contacted, one can afford ~ 1 mA current flow in a region only ~ 20 nm wide and ~ 2 nm tall. This is significant considering that a 50 nm thick copper film needs to be ~ 2 μm wide in order to carry the same current without breakdown by electro-migration. Thus, short carbon nanotubes are promising interconnect materials with optimum current carrying ability, low power dissipation and superior chemical stability.

With individual semiconducting SWCNTs, we have formed nanotube FETs with L~50 nm long channels using the shadow evaporation technique. Note that the gate dielectric in our current work is ~10 nm thick $SiO_2$ in a back-gate configuration. Although even shorter FETs can be fabricated using our method, we have limited L~>50 nm to avoid short channel effects. Shown in Fig.4 is a L~50 nm SWCNT-FET ($d$~2 nm) with on- and off-current ratio $I_{on}/I_{off}$ ~$10^3$ at $V_{DS}$=0.3 V, subthreshold swing of ~300 mV/decade (Fig.4b), and transconductance $(dI_{DS}/dV_G)_{max}$ ~ 7 µS. The most notable property of the device is that high current of ~20 µA can be reached at a low bias voltage of $V_{DS}$~0.4 V (Fig.4c). In comparison, similar currents can be reached only under much higher bias of $V_{DS}$ ~2 V for L~3 µm long SWCNT FETs (9). This suggests that the L~50 nm SWCNT FETs are significantly more ballistic than the long channel devices under high bias and current conditions (10). Such transistors are appealing for ultra-fast electronics since the on-state current is directly proportional to the speed of a transistor. Further channel length scaling to the L ~ 5 nm < $l_{op}$ scale is needed to approach the ultimate ballistic transport limit for CNT-FETs.

In summary, we have obtained large numbers of ultra-short carbon nanotube electrical devices without using sophisticated electron-beam lithography tools. Our simple shadow evaporation method allows for length scaling of SWCNT electronics down to 10 nm and the elucidation of transport properties at this length scale. While short metallic nanotubes are nearly ballistic conductors useful for future interconnects, short semiconducting nanotubes can be exploited for near ballistic transistors for high current operations. In addition, the simple fabrication technique can easily be applied to other materials for obtaining ultra-miniaturized devices including nanowires and nanorods.  Thus, highly scaled devices based on chemically derived nanomaterials can now be fabricated in any laboratory with access to simple photolithography tools.

**Acknowledgement.**  We thank Marco Rolandi for assistance with an experiment. This work was supported by MARCO MSD Focus Center, Stanford INMP, DARPA Moletronics, SRC/AMD, DARPA MTO, a Packard Fellowship, NSF Network for Computational Nanotechnology, and an SRC Peter Verhofstadt Graduate Fellowship (A. J.).

**Figure Captions**

**Figure 1.** Current ($I_{DS}$) vs. bias voltage ($V_{DS}$) characteristics of a ~ 1.1 μm long single-walled carbon nanotube. The current exhibits saturation under high biases due to scattering by optical phonons with a short mean free path of $l_{op}$~ 10-15 nm. Inset: AFM image of the device (SWCNT diameter d~2.5 nm). The line-like structures near the top and bottom of the image correspond to the edges of the S and D Pd electrodes respectively. Note that the typical diameters of nanotubes used in the current work are in the range of 1.5-3 nm.

**Figure 2.** A ~10 nm long metallic nanotube device formed by an angle evaporation method. (a) A schematic illustration of the device formation process. A pre-formed D electrode is used to mask metal deposition directed at an angle θ with respect to the substrate normal. (b) SEM image of a L~15 nm long fabricated SWCNT device. The dark line-like region between the S and D electrodes is ~ 15 nm in width and is bridged by a SWCNT. Note that we have not used carbon nanotube AFM tips extensively in the current work to characterize the narrow gaps for the following reasons. Multi-walled CNT tips have diameters on the order of 10-30 nm, too large for imaging the narrow gaps. SWCNT tips can be small (down to ~1 nm in diameter) but are typically short (~20 nm, for mechanical stiffness) and not suitable for imaging the gaps due to the 30-50 nm tall D-electrodes. Our imaging attempts with SWCNT tips encountered the problem of the short tips being incapable of reaching down into the gaps. (c) $I_{DS}$-$V_{DS}$ characteristics

of the L~15 nm nanotube (d ~ 2 nm). The lines are drawn to show the slopes of the curve in different bias regimes.

**Figure 3.** Multiple (8) ultra-short (L~15 nm) SWCNTs electrically connected in parallel to carry near 1mA current. (a) AFM image of a device comprised of eight short (L ~ 15 nm) tube-segments connected in parallel. The eight short-tube devices were formed on a single nanotube (d~2 nm) by the following process. A set of parallel metal electrodes (250 nm in width) was first formed on top of the nanotube (electron beam lithography was involved for this experiment). These electrodes were then all used as shadow masks during angle evaporation for the second metal deposition step. Each electrode cast a shadow (dark lines in image) bridged by the nanotube. The multiple shadow devices were then connected in parallel in a comb-like configuration, as drawn in the figure. The bottom panel shows a schematic side-view structure (left) and top-view AFM image (right) for one of the eight devices. (b) $I_{DS}$-$V_{DS}$ characteristic of the 8 tube-device, showing little current saturation at high biases and a current carrying capability near 1mA.

**Figure 4.** A L ~ 50 nm long channel nanotube field-effect transistor formed by the shadow evaporation method without electron beam lithography. (a) AFM tapping-mode phase image of a device with a SWCNT (d ~ 2 nm) bridging source and drain electrodes (drain formed by the shadow method). (b) $I_{DS}$-$V_{GS}$ curves recorded at $V_{DS}$= -0.1, -0.2 and -0.3 V respectively. (c) $I_{DS}$-$V_{DS}$ curves recorded at various gate voltages as indicated for the 50 nm SWCNT FET. Note that metal deposition by electron beam evaporation is largely directional but diffusive to a small degree. Small numbers of metal atoms may deposit onto the SWCNT in the shadow region, which could then alter the transistor characteristics. We have observed such an effect when attempting channel length scaling down to 10 nm or below. Also note that it is difficult to absolutely ensure that there is a single SWCNT in a device. We used AFM or SEM to ensure that there is only a single connection between the S and D. We also often carried out electrical breakdown of the devices at the end of the measurements. Occasional devices exhibiting two current drops (corresponding to breakdown of a two-tube raft or double-walled CNT) are excluded and not presented here.

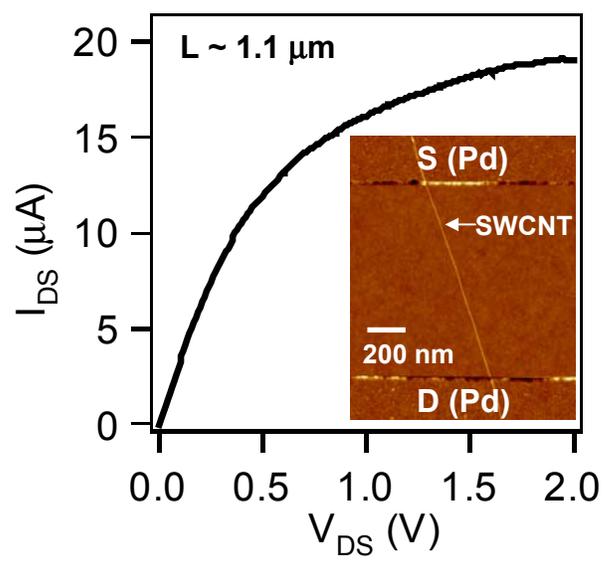

Figure 1

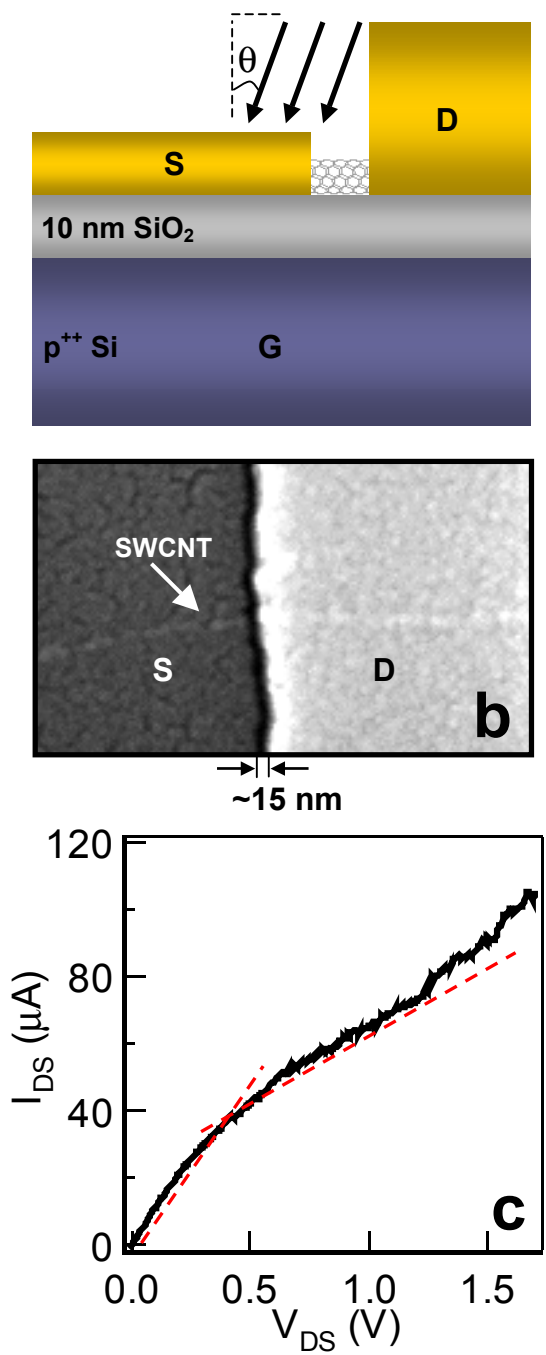

Figure 2

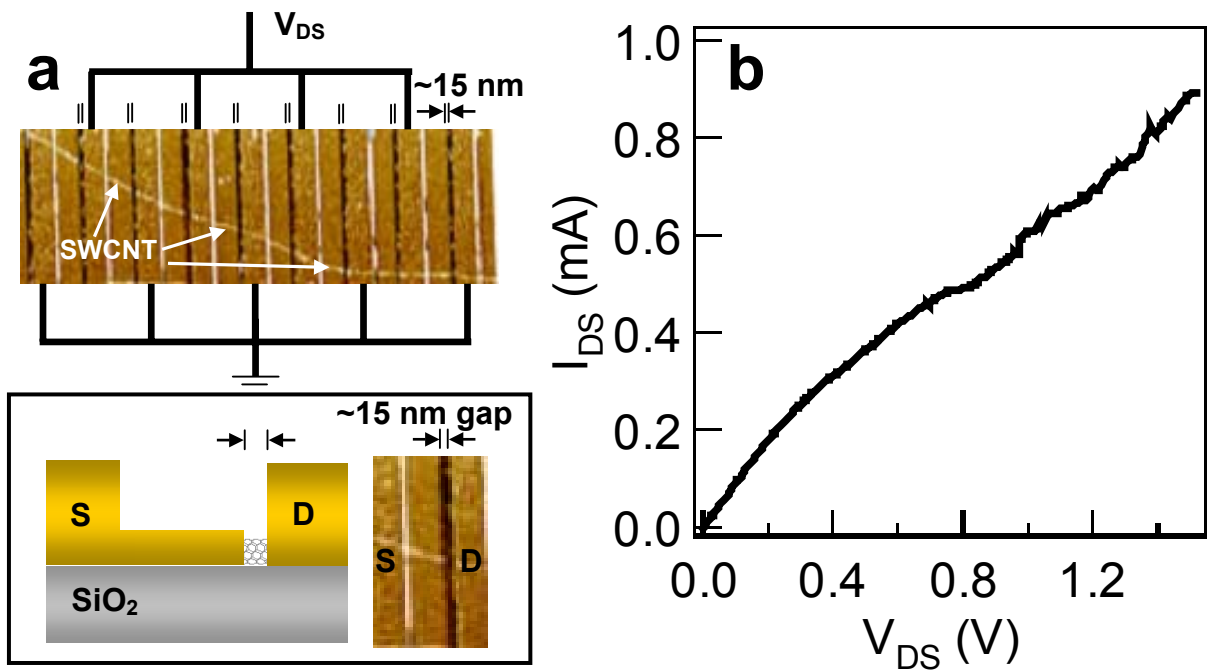

Figure 3

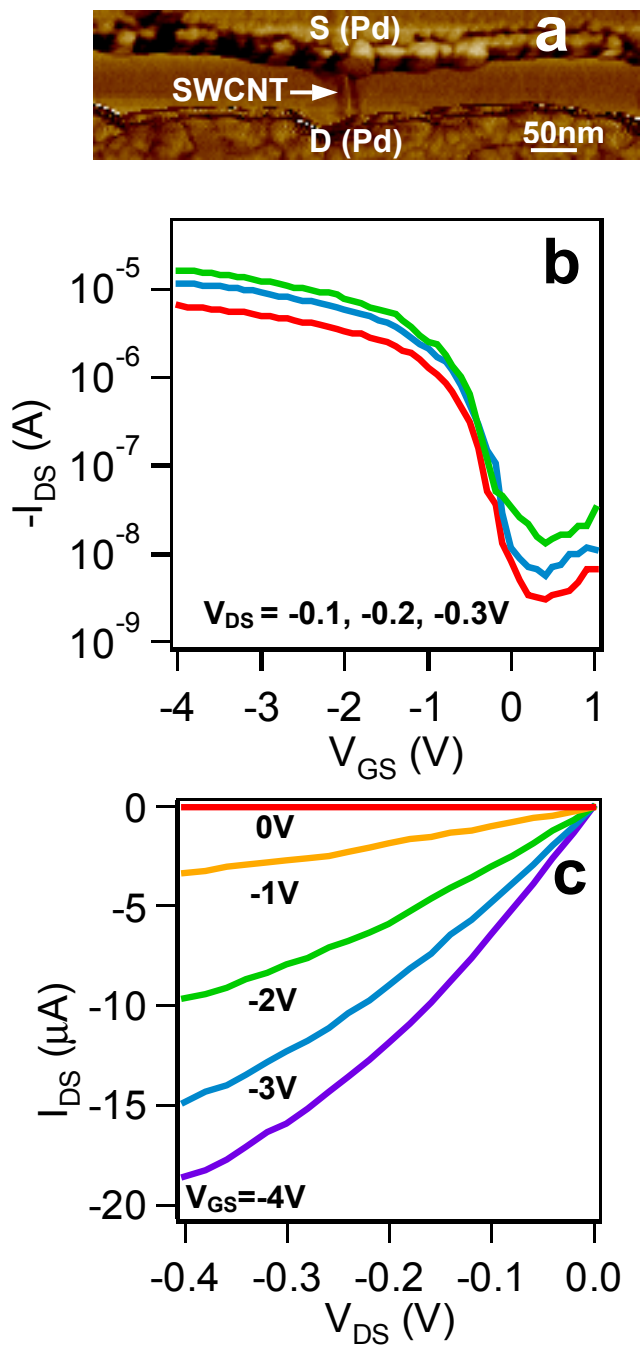

Figure 4